\documentclass[prl,reprint,superscriptaddress]{revtex4-2}
\usepackage{soul}
\usepackage{color}
\usepackage{graphicx}
\usepackage{subcaption}
\usepackage{amsmath}
\usepackage{array}

\usepackage{float}
\usepackage{lineno}
\usepackage{ulem}
\usepackage{siunitx}
\usepackage[bottom]{footmisc}
\usepackage{ragged2e}
\usepackage[dvipsnames]{xcolor}
\definecolor{apsblue}{HTML}{2E3092}
\usepackage[colorlinks=true,
            linkcolor=apsblue,
            citecolor=apsblue,
            urlcolor=apsblue]{hyperref}
\newcommand{\rmax}{r_{\rm max}}
\begin{document}


\title{Strong Constraints on Higgsino Dark Matter from Solar Capture}


\def\there{Work}
\def\here{Home}

\def\um{William I. Fine Theoretical Physics Institute,University of Minnesota, MN 55455, United States}
\def\cern{Theoretical Physics Department, CERN, 1 Esplanade des Particules, CH-1211 Geneva 23, Switzerland}
\def\udelbartol{Department of Physics and Astronomy, University of Delaware and the Bartol Research Institute, Newark, DE 19716, USA}

\author{Maxim Pospelov}\affiliation{\um}\affiliation{\cern}
\author{Harikrishnan Ramani}\affiliation{\udelbartol}

\date{\today}
\begin{abstract}
Higgsino dark matter remains one of the most convincing dark matter candidates. It can exist in a ``quasi-Dirac" regime, when the splitting between two Majorana components is on the order of $O(\rm few~\,MeV)$ or less. Owing to the strong gravitational pull of the Sun, the DM particles accelerate to up to $\sim 1400$\,km/sec in the solar core, which allows for inelastic scattering on heavy elements and very effective capture onto solar-bound trajectories. We evaluate the expected flux of the neutrinos from the expected $\chi\chi \to W^+W^-,ZZ$ annihilation for the cosmologically preferred mass, $m_\chi \simeq 1.08$\,TeV while keeping the mass splitting $\delta$ as a free parameter. Confronting it with the non-observation of high-energy neutrinos from the Sun by the IceCube experiment, we obtain a robust limit, $\delta>566$\,keV. These limits exclude the interpretation of the recent LZ event in terms of endothermic inelastic scattering of Higgsino dark matter on xenon nuclei.
\end{abstract}

\maketitle
\textit{\textbf{Introduction.}}
Electroweak multiplets rank among the best-motivated dark matter candidates. Their construction is minimal: dark matter couples to the visible sector through Standard Model gauge interactions alone, with no additional mediators. Such states arise naturally in extensions of the Standard Model, most notably supersymmetry, where the Higgsino and Wino furnish canonical examples. Moreover, for masses at the poorly explored TeV scale, electroweak annihilation cross sections yield a thermal relic abundance matching the observed dark matter density; the celebrated ``WIMP miracle". 

It is remarkable that the simplest electroweak WIMPs have survived decades of experimental scrutiny. Colliders constrain these multiplets only up to a few hundred GeV, far below their thermal masses; reaching them will likely require a muon collider~\cite{Han:2020uak,Capdevilla:2021fmj,Bottaro:2022one} or a $100\,\mathrm{TeV}$ hadron machine~\cite{Saito:2019rtg}. Indirect detection places stringent constraints on the smallest real multiplets (hypercharge $Y=0$)~\cite{Baumgart:2025dov,Safdi:2025sfs}, whereas the prospects for complex WIMPs ($Y \neq 0$) are less favorable~\cite{Abe:2025lci}. Direct detection would have long excluded complex WIMPs through spin-independent tree-level $Z$ exchange; they survive only in models where Majorana masses, generated after electroweak symmetry breaking, split the Dirac state into two Majorana eigenstates and render the $Z$-mediated scattering entirely inelastic. This mechanism arises naturally in supersymmetry, where the Higgsino, a complex electroweak doublet, acquires Majorana mass splitting $\delta$ from Wino and Bino mixing, making it notoriously difficult to detect.

The mass splitting $\delta$ is thus the only free parameter for a given electroweak WIMP. This parameter does not affect the abundance of the Higgsino DM, and following detailed calculations in the literature, we shall assume that the mass leading to the saturation of the primordial dark matter abundance is $m_\chi \simeq 1.08\,\rm TeV$. The splitting $\delta$ has the most profound effect on the Higgsino-nucleus scattering cross section at typical dark matter velocities. Indeed, for $\delta \to 0$, the cross section is $\sim$ ten orders of magnitude larger than for $\delta$ in excess of several MeV, where the tree level $Z$ exchange channel completely shuts off. In the intermediate regime, for scattering off a nucleus of mass $m_A$, inelastic kinematics requires $\delta < \frac{1}{2}\mu_A v^2$, where $\mu_A$ is the reduced mass of the dark matter--nucleus system and $v$ is the dark matter velocity. Xenon-based experiments are therefore blind to $\delta \gtrsim 368~\mathrm{keV}$, a ceiling set by the fastest dark matter allowed by the Milky Way escape velocity. To add further intrigue to the Higgsino parameter space, the LZ collaboration recently reported one event in the high recoil region~\cite{LZ:2026extended} consistent with a TeV dark matter event with splitting in the $\{340~\textrm{keV},360~\textrm{keV}\}$ window.   

Proposals to probe larger splittings rely on heavier target elements~\cite{Eby:2019mgs, Graham:2024syw, Graham:2026ivn,Alloni:2026xdf} or on components of the dark matter population that exceed the galactic escape velocity~\cite{Graham:2024syw, Graham:2026ivn}. The Sun is a natural environment that combines both advantages: heavy elements such as iron and beyond are present, and infalling dark matter is accelerated to the solar escape velocity, $v_\mathrm{esc} \gtrsim 600~\mathrm{km/s}$ at the surface and exceeding $1300~\mathrm{km/s}$ in the core. Solar scattering is therefore sensitive to splittings well beyond terrestrial reach, provided one can identify a probe of this population. If dark matter scatters in the Sun and loses sufficient energy, it becomes gravitationally captured; subsequent annihilations copiously produce TeV-scale neutrinos detectable at neutrino telescopes such as IceCube. Indeed, IceCube sets stringent limits on a variety of dark matter models through this channel~\cite{IceCube:2025fcu}. This has been explored in the past to constrain inelastic dark matter explanations~\cite{Nussinov:2009ft,Menon:2009qj,Tucker-Smith:2001myb} to the DAMA anomaly and for spin-dependent elastic scattering for the mixed Higgsino-gaugino scenario~\cite{Krall:2017xij}. To our knowledge, however, this technique has not been applied to inelastic electroweak dark matter, of which Higgsino dark matter is the prime example.

\textit{\textbf{Solar Model.}}
We model the dark matter distribution to have an asymptotic density $\rho_{\rm DM} = 0.3~\mathrm{GeV\,cm^{-3}}$ and a truncated Maxwell--Boltzmann velocity distribution in the Galactic frame,
\begin{equation}
    f_{\rm mb}(\vec{v}) = \frac{1}{N_{\rm esc}} \frac{1}{(\pi v_0^2)^{3/2}} \, e^{-v^2/v_0^2} \, \Theta(v_{\rm esc}^{\rm gal} - v),
\end{equation}
with dispersion parameter $v_0 = 220~\mathrm{km\,s^{-1}}$ and Galactic escape velocity $v_{\rm esc}^{\rm gal} = 544~\mathrm{km\,s^{-1}}$, and the normalization $N_{\rm esc}$ ensures that $\int d^3v\, f_{\rm mb}(\vec{v}) = 1$.
The distribution in the solar frame follows from the Galilean boost
\begin{equation}
    f_u(\vec{u}) = f_{\rm mb}(\vec{u} + \vec{v}_\odot),
\end{equation}
with $v_\odot \simeq 232~\mathrm{km\,s^{-1}}$ the solar velocity with respect to the Galactic rest frame.

We use Liouville's theorem to determine the dark matter flux at an arbitrary radius $r$ inside the Sun. Although the solar motion renders $f_u(\vec{u})$ anisotropic, the spherical symmetry of the solar potential ensures that capture depends only on the angle-averaged speed distribution, $F(u) \equiv u^2 \int d\Omega_u\, f_u(\vec{u})$. Energy conservation relates the local speed to the asymptotic speed via $w(r,u) = \sqrt{u^2 + v_{\rm esc}^2(r)}$, with $v_{\rm esc}(r)$ the local solar escape velocity, and conservation of phase-space density then gives the differential flux
\begin{equation}
    \frac{d\Phi}{du} = \frac{\rho_{\rm DM}}{m_\chi}\, \frac{F(u)}{u}\, w^2(r,u)\,.
\end{equation}

Next we use the BS05(AGS,OP)~\cite{Bahcall:2004pz} for the density profile of the Sun. We make a justifiable assumption that heavy elements have the same density profile up to an overall normalization. We can use the total density profile to also estimate $v_{\rm esc}(r)$ in the Sun. 

\begin{figure}
    \centering
    \includegraphics[width=0.99\linewidth]{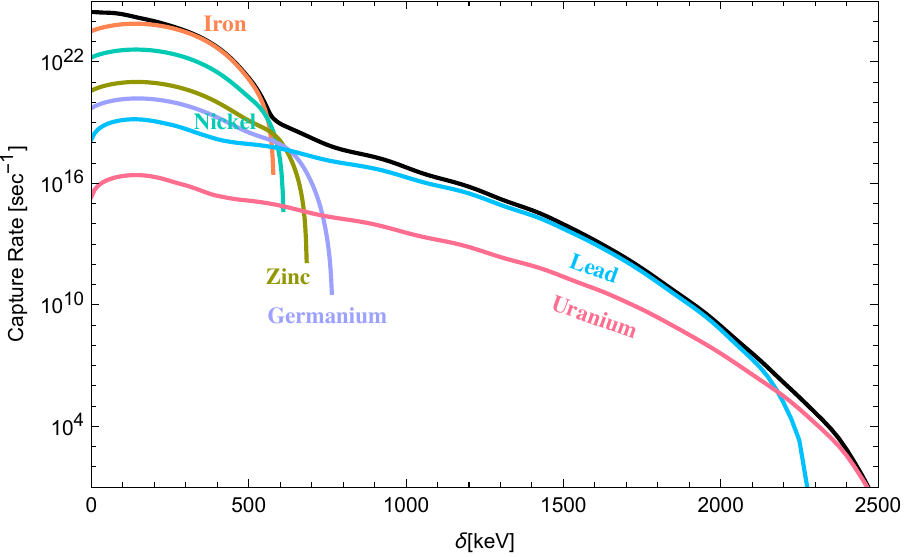}
    \caption{Capture rate for the 1.08 TeV Electroweak doublet (Higgsinos) with different elements in colors and total (black) as a function of the neutral mass splitting.}
    \label{fig:ratecalc}
\end{figure}

\textit{\textbf{Capture.}}
We define a capture process as one where the in-fall dark matter loses enough energy to fall below the escape velocity at the scattering point. The differential cross-section is given by~\cite{Essig:2007az,Goodman:1984dc}
\begin{align}
\frac{d\sigma_{\chi N}}{dE_R}\big(w,E_R\big)
   \;&=\; \frac{G_F^{2}\, m_A}{4\pi\, w^{2}}\,
   \Big[\,N - \big(1-4\sin^{2}\theta_W\big)Z\,\Big]^{2}\, \nonumber \\
  &\times  F_A^{2}\!\left(\sqrt{2m_A E_R}\right),
\label{eq:dsdER}
\end{align}
In the limit of $\delta\to 0$, this gives a total cross section of 
$\sigma_{\chi N} = (2 \pi)^{-1}G_F^2 \mu_A^2\Big[\,N - \big(1-4\sin^{2}\theta_W\big)Z\,\Big]^{2}$.
Note that this cross-section is a factor of 4 larger than the one that has propagated in literature~\cite{Cirelli:2005uq,Nagata:2014wma,Bramante:2016rdh,Krall:2017xij,Graham:2024syw} which would correspond to scattering with a left-handed electroweak state, rather than a vector-like Higgsino. 
Here $G_F$ is the Fermi constant, $m_A$ is the mass of the nucleus, $E_R$ is the energy recoil, $Z$ and $N$ are the number of protons and neutrons respectively and $F_A$ is the nuclear from factor. Following the dark matter literature, we assume it to be the Helm form-factor. The kinematic limits of $E_R$ are given by, $E_\pm = \frac{\mu_A^{2}\,\big(w \pm w'\big)^{2}}{2\,m_A}\,$ with
$w' = \sqrt{\,w^{2} - 2\delta/\mu_A\,}\,$. Thus, the total capture rate is given by,
\begin{align}
\mathcal{C} \;=\; \sum_A \int_0^{R_\odot}\! 4\pi r^2\, dr\; n_A(r)
\int_0^{u_{\max}}\! du\;\frac{d\Phi}{du}(r,u) \nonumber \\
\times \int_{E_R^{\min}}^{E_{+}}\! dE_R\;
\frac{d\sigma_{\chi N_A}}{dE_R}\big(w(r,u),E_R\big)\,,
\label{eq:capture}
\end{align}
Here, 
\begin{equation}
E_R^{\min} = \max\!\left[\,E_-\,,\; \tfrac{1}{2} m_\chi u^{2} - \delta\,\right]
\end{equation}
with the second term ensuring that the dark matter loses at least $\tfrac{1}{2} m_\chi u^{2}$ of kinetic energy in order to get captured and $u_{\rm max}$ is the largest asymptotic speed for which capture on species
$A$ at radius $r$ remains kinematically allowed and $n_A(r)$ is the number density of species $A$ at radius $r$.
This rate is plotted as a function of $\delta$ in Fig.~\ref{fig:ratecalc}. 

\begin{figure}
    \centering
    \includegraphics[width=0.95\linewidth]{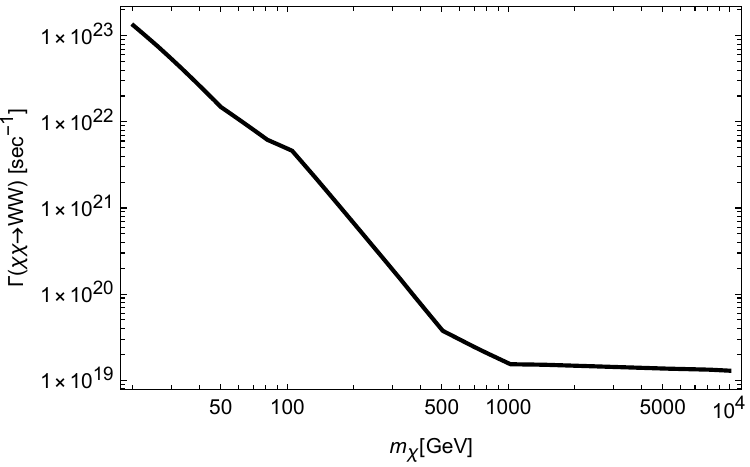}
    \caption{Inferred IceCube limit on dark matter annihilations to $W^+W^-$ in the Sun.}
    \label{fig:annrate}
\end{figure}

\textit{\textbf{Derived Limit from IceCube.}}
This capture rate also acts as an upper-bound on the annihilation rate to neutrinos, with the bound saturating at equilibrium between the two processes. To this end, we estimate the constraint on neutrino annihilation from the IceCube experiment that sets strong constraints on $\nu_\mu$ and $\bar\nu_\mu$ fluxes at Earth correlated with the direction to the Sun. We use the limits from 10 year data published by IceCube~\cite{IceCube:2025fcu} where equilibration between annihilation and capture is assumed. Using the published limits on spin-dependent and spin-independent cross-section, we reverse-engineer the relevant capture rate that is constrained. This procedure has the added advantage that it avoids the need to redo neutrino propagation in the Sun, since that is already included in the above analysis by the IceCube collaboration. Using the spin-dependent analysis in ~\cite{IceCube:2025fcu}, and the capture rates~\cite{Jungman:1995df} as used in ~\cite{IceCube:2025fcu}, we obtain an annihilation rate to $W^+W^-$ that is plotted in Fig.~\ref{fig:annrate}. Specifically for the Higgsino mass of 1.08 TeV, we obtain a constraint  
\begin{equation}\Gamma_\textrm{ann}^\textrm{lim}(\chi \chi \rightarrow W^+W^-)=1.5 \times 10^{19}~\textrm{sec}^{-1}.
\label{eq:annh}
\end{equation}
Notice that $W^+W^-$ is of course the dominant mode of the Higgsino annihilation. 

\textit{\textbf{Evolution of captured Dark Matter.}}

\begin{figure}
    \centering
    \includegraphics[width=0.99\linewidth]{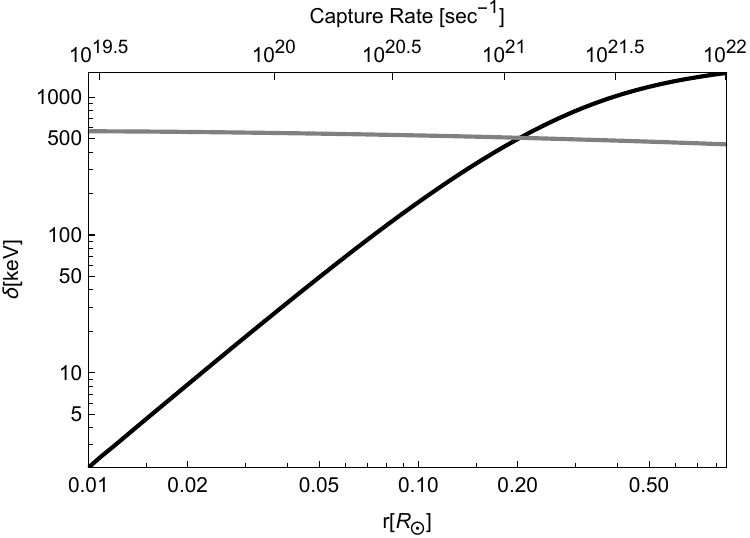}
    \caption{Black: $\delta$ as a function of the stall radius $r$ in uranium.
The top axis gives the capture rate $\mathcal{C}$ required for an
annihilation rate equal to the current IceCube constraint. Gray: $\delta$
versus the capture rate $\mathcal{C}$, the inverse of the black curve in
Fig.~\ref{fig:ratecalc}. Tree-level processes alone therefore constrain
$\delta$ below the intersection of the two curves, i.e. $\approx 506$ keV}
    \label{fig:uranium}
\end{figure}

In order to set limits we need the annihilation rate of the dark matter in the Sun, which depends on its radial profile after capture. With the simple assumption that all the captured dark matter falls within a radius $R$ in fast time scales compared to the age of the Sun $t_\odot$, we can estimate the time for equilibration $\tau_{\rm eq}$ between capture and annihilation   via
\begin{equation}
\frac{\mathcal{C}\tau_{\rm eq}}{\frac{4}{3}\pi R^3 } \times \mathcal{C}\tau_{\rm eq} \langle \sigma v \rangle _{\rm ann} =\mathcal{C}
\end{equation}
which gives,
\begin{align}
\tau_{\rm eq}=&4.5\times 10^9~\textrm{year} \times \left(\frac{\mathcal{C}}{3\times 10^{19}\text{sec}^{-1}}\right)^{-\frac{1}{2}}\left(\frac{R}{0.017 R_\odot}\right)^\frac{3}{2}\nonumber \\ &\quad \times \left(\frac{1.3 \times 10^{-26} \textrm{cm}^3\textrm{sec}^{-1}}{\langle\sigma v\rangle_{\rm ann}}\right)^{\frac{1}{2}}.
\label{eq:teq}
\end{align}
Here, we use the reference annihilation cross-section ${\langle\sigma v\rangle_{\rm ann}}$ from \cite{Beneke:2014hja}  and it is given by ${\langle\sigma v\rangle_{\rm ann}}=1.3 \times 10^{-26} \textrm{cm}^3\textrm{sec}^{-1}$. 
Since the Sun is over 4.5 billion years old, dynamical equilibrium can be safely assumed only if the typical radius of Higgsino sphere falls below $0.017 R_\odot$. In general we obtain
\begin{equation}
\Gamma_{\rm ann}(t)= \frac{\mathcal{C}}{2} \tanh^2 \left(\frac{t}{\tau_{\rm eq}}\right)
\label{eq:annrate}
\end{equation}
Thus, in order for $\Gamma_{\rm ann}$ to exceed $\Gamma^{\rm lim}_{\rm ann}$ in Eq.~\ref{eq:annh}, either there needs to be subsequent thermalization that further concentrates the dark matter at the center or $\mathcal{C}$ should greatly exceed $\Gamma^{\rm lim}_{\rm ann}$. The top and bottom $x$-axes of Fig.~\ref{fig:uranium} correspond to the capture rate and radius respectively for $\tau_{\rm eq}=4.5 \times 10^9 \textrm{year}$. 

The dark matter is mainly captured in the extremely dense core of the Sun i.e. within $0.1 R_\odot$. However, it can still make long orbits and spend significant time outside this radius. Once captured, the dark matter slows down by scattering with increasingly heavier elements. With uranium, for a given splitting $\delta$, the scattering stops when the dark matter falls below a speed $v_{U,\delta}$ given by $\frac{1}{2} \mu_U v_{U,\delta}^2=\delta$. This compacts the dark matter to a radius $r$ determined from the relation 
\begin{equation}
v_{\rm esc}(0)^2-v_{\rm esc}(r)^2=v^2_{U,\delta}.
\end{equation}

We plot $\delta$ as a function of the stall radius $r$ in black in Fig.~\ref{fig:uranium}  for uranium. The top $x$-axis also has the capture rate that would be required so as to achieve an annihilation rate of the one constrained by IceCube today. Now, in gray, we plot $\delta$ as a function of the capture rate $\mathcal{C}$. This is the inverse of the black line in Fig.~\ref{fig:ratecalc}, and very small variations in $\delta$ result in large range for $\mathcal{C}$. Hence, just tree-level processes constrain $\delta$ which are smaller than the intersection of gray and black lines. This corresponds to $\delta\approx506~\textrm{keV}$. This is by itself a 40\% improvement over existing bounds from direct detection~\cite{LZ:2026extended}. 


%
%
After the tree-level process stalls, the particle is on a nearly radial orbit
whose turning point $\rmax$ 
is approximately $\rmax^{(0)}\simeq0.20\,R_\odot$ for $500\textrm{keV}\lesssim\delta\lesssim580\,$keV.  Further
energy loss proceeds only through the loop level elastic
channel. The spin-independent cross-section $\sigma^{\rm SI}_{\rm loop}$, famously suffers from a destructive interference between diagrams and can be several orders of magnitude below naive estimates. We use the calculated values in literature~\cite{Chen:2019gtm} with the central value $4\times 10^{-50}~\textrm{cm}^2$, the upper boundary is $3.16\times 10^{-49}~\textrm{cm}^2$ and the lower  boundary is consistent with zero. The spin-dependent part derived in \cite{Hisano:2011cs} is approximately $5\times10^{-47}~\textrm{cm}^2$. While it is usually neglected in favor of the spin-independent cross-section for direct detection, it is more relevant here because of the dominance of hydrogen in the sun and the perverse cancellation in the SI cross-section.

\paragraph{Orbit.}
The radial orbit with turning point $\rmax$ has local speed

 $ v(r;\rmax)=\sqrt{\,v_{\rm esc}^2(r)-v_{\rm esc}^2(\rmax)\,}$ and orbital time 
 \begin{align}
  T(\rmax)=4\!\int_0^{\rmax}\!\frac{dr}{v(r;\rmax)}
  \end{align}
%

\paragraph{Slowing cross section.}
For target species $A$ the elastic differential cross section is
\begin{equation}
  \frac{d\sigma_A}{dq^2}
  = \frac{A^2\,\sigma^{\rm SI}_{\rm loop}F_A^2(q)+\sigma^{\rm SD}_{\rm loop}}{4\mu_n^2 v^2}\,,
  \qquad
  q_{\max}=2\mu_A v ,
  \label{eq:dsdq2}
\end{equation}
with $\mu_n$ ($\mu_A$) the dark matter--nucleon (--nucleus) reduced mass and
$F_A$ the Helm form factor.  In practice, we only include the contribution from Hydrogen in $\sigma_{\rm loop}^{\rm SD}$, for which the form factor can be safely put to one. A recoil $E_R=q^2/2m_A$ removes a fraction
$E_R/E_\chi$ of the kinetic energy $E_\chi=\tfrac12 m_\chi v^2$, so the
relevant weighting is the slowing cross section
\begin{equation}
  \sigma^{\rm slow}_A(v)
  \;\equiv\;\int_0^{q^2_{\max}}\!\frac{E_R}{E_\chi}\,
      \frac{d\sigma_A}{dq^2}\,dq^2
  \label{eq:sigslow}
\end{equation}
%

%
%

\paragraph{Evolution of the turning point.}
Let $E=\tfrac12 v^2+\Phi(r)$ be the energy per unit dark matter mass, with
$\Phi(r)=-\tfrac12 v_{\rm esc}^2(r)$, so that the turning point is fixed by
$E=\Phi(r_{\rm max})$. At radius $r$ this energy is drained at the rate
\begin{equation}
  \left.\frac{dE}{dt}\right|_r
  = -\tfrac12 v^2(r)\sum_A n_A(r)\,v(r)\,\sigma_A^{\rm slow}\big(v(r)\big),
\end{equation}
where $v \equiv v(r;r_{\rm max})$. Averaging over one radial period using
$dt = dr/v$ and Eq.~(12),
\begin{align}
  \Big\langle \frac{dE}{dt}\Big\rangle
  = -\frac{2}{T(r_{\rm max})}\int_0^{r_{\rm max}}\!\!\sum_A n_A(r)\,
    \sigma_A^{\rm slow}\nonumber \\ \times \big(v(r;r_{\rm max})\big)\,v^2(r;r_{\rm max})\,dr ,
\end{align}
where $\langle\cdot\rangle$ denotes the average over a radial orbit. The turning
point then responds as $\dot E = g(r_{\rm max})\,\dot r_{\rm max}$, with
$g(r)=GM(r)/r^2$ and $M(r)$ the mass enclosed inside radius $r$, so that
\begin{equation}
  \frac{d r_{\mathrm{max}}}{dt}
  = \frac{1}{g(r_{\mathrm{max}})}\Big\langle \frac{dE}{dt}\Big\rangle .
  \label{eq:master}
\end{equation}
\begin{figure*}[htbp]
    \centering
    \begin{subfigure}[b]{0.49\textwidth}
        \centering        \includegraphics[width=\linewidth]{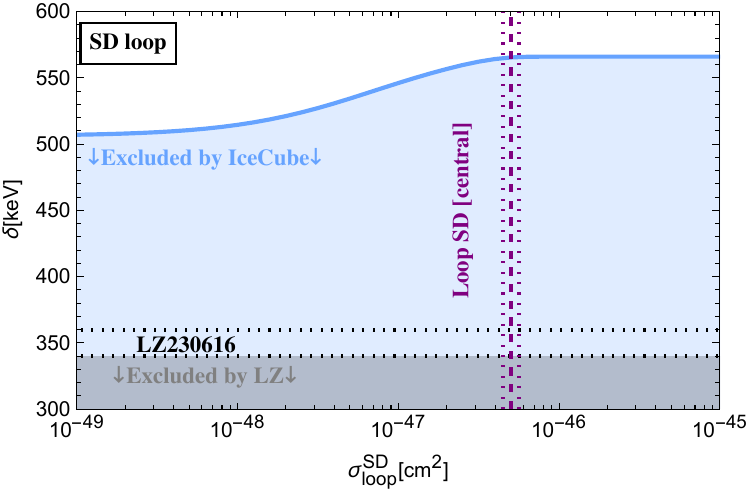}   
        \label{fig:panel_a}
    \end{subfigure}
    \begin{subfigure}[b]{0.49\textwidth}
        \centering
\includegraphics[width=\linewidth]{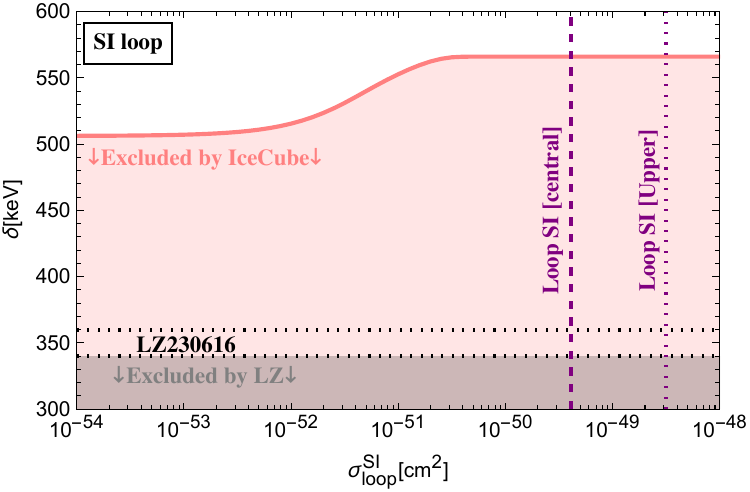}
    \label{fig:panel_b}
    \end{subfigure}
    \caption{Constraints on the  mass splitting $\delta$ for Higgsino dark matter as a function of the loop cross-section. for SD cross-section (\textbf{Left}), the constraint is robustly 566 keV. For SI loop cross-section (\textbf{Right}), constraints range from $506$ keV for negligible loop cross-section to $566$ keV for cross-sections that exceed $\approx 10^{-50}~\textrm{cm}^2$. The central loop cross-section from~\cite{Chen:2019gtm} as well as the upper bound are shown. The lower bound is consistent with zero. The best fit region for the single event from~\cite{LZ:2026extended} as well as the associated exclusion are also shown.}
    \label{fig:finalexc}
\end{figure*}

This process stalls when the dark matter thermalizes with the SM plasma. The thermalization radius is given by
\begin{equation}R_{\rm therm}=\frac{3^{5/6} \sqrt{\frac{T_{\rm core}}{G m_\chi \rho_{\rm core}}}}{2 \sqrt[6]{2} \sqrt[3]{\pi }}\approx 0.0037 R_\odot\times \sqrt{\frac{1\textrm{ TeV}}{m_\chi}},
\end{equation}
which is comfortably smaller than the $0.017 R_\odot$ we obtained in Eq.~(\ref{eq:teq}). 

We can obtain $\rmax(t=t_\odot)$ by solving the differential equation in Eqn.~\ref{eq:master}. Substituting this into Eqn.~\ref{eq:teq} and asking for the resultant annihilation rate in Eqn.~\ref{eq:annrate} to exceed the IceCube limit in Eqn.~\ref{eq:annh}, we obtain a limit on the splitting $\delta$ as a function of the loop cross-section $\sigma_{\rm loop}$, and we distinguish spin-independent (SI) and spin-dependent (SD) scattering. We plot the resultant exclusion in Fig.~\ref{fig:finalexc}. We see broadly two regimes. The large cross-section regime where thermalization happens within $t_\odot$ corresponds to a limit of $\delta=566~\textrm{keV}$. This can be obtained from Fig.~\ref{fig:ratecalc} by equating the captured rate with the rate limited by IceCube. As we dial the SI cross-section down, we encounter a bridge (or bend), which finally asymptotes to $\delta\approx506~\textrm{keV}$ for small-enough cross-sections. This asymptote corresponds to the stall radius for Uranium obtained earlier. The SD-induced energy loss has a very similar bend at larger cross section values. It is important to reiterate that while the SI cross-section has large error bars and can be very suppressed, the SD value is more robustly predicted and is not subject to very strong cancellations. Taking the predictions from Ref.\,\cite{Hisano:2011cs}, we see that the SD scattering of Higgsino on hydrogen is enough to equilibrate capture and annihilation, so that the stronger limit of 566\,keV should apply. 
Regardless of the post-capture energy loss treatment, the best-fit splitting from the LZ event LZ230616 is robustly ruled out by this analysis. The constraints are also an improvement over the excluded region from LZ.

 \textit{\textbf{Conclusions.}} The Sun accelerates infalling dark matter to speeds
unattainable in any terrestrial experiment, and that single fact makes it the
most sensitive available probe of the inelastic Higgsino. For
$m_\chi = 1.08$ TeV we find that the non-observation of high-energy neutrinos
from the Sun in ten years of IceCube data excludes neutral mass splittings
$\delta \lesssim 506$ keV. This bound rests on tree-level $Z$ exchange alone:
dark matter up-scatters
repeatedly off uranium until the process shuts off kinematically near
$0.2\,R_\odot$, and the annihilation rate that follows already exceeds
$\Gamma_{\rm ann}^{\rm lim}$. It reaches a factor of $1.4$ beyond terrestrial
direct detection, which is blind above $\delta \simeq 368$ keV for any cross
section.

   For the predicted loop-induced elastic cross section values (notwithstanding uncertainties of $\sigma^{\rm SI}$) the captured population
continues to sink, and the bound strengthens to $\delta \lesssim 566$ keV. 
What
makes this second number useful is how little it depends on
$\sigma_{\rm loop}$. 
The limit moves by
only $\approx 60$ keV across several decades of cross-section in Fig.~\ref{fig:finalexc}. 
The
immediate consequence is that the recent LZ event~\cite{LZ:2026extended} (that  has already invited several explanations \cite{Su:2026LZ,Fan:2026LZ,Freese:2026LZ,Wu:2026LZ,Lou:2026LZ} including some based on Higgsino inelastic scattering) cannot be from the inelastic Higgsino
scattering: fitting the observed recoil requires $\delta \simeq 340$--$360$ keV. At this splitting, as shown in this work, the IceCube limits will be violated by $O(10^3)$ in the annihilation rate,
and that range is excluded for every value of $\sigma_{\rm loop}$, including zero.

While we find that the Higgsino-induced interpretation of the LZ event is excluded by the high energy neutrinos from the Sun, other models of inelastic dark matter can be easily compatible with the result of~\cite{LZ:2026extended}. For example, inelastic models based on dark photon mediation \cite{Arkani-Hamed:2008hhe,Batell:2009vb,Finkbeiner:2009mi} can be invoked. Indeed, if the mass of dark photons is sub-GeV, the process $\chi\chi\to A'A'$ annihilation will not produce prompt neutrinos, and for a large range of the kinetic mixing parameter the solar capture will not produce significant constraints. 

\textit{\textbf{Acknowledgements.}}
HR is supported in part by NSF
Grant No. PHY-2515007, The University
of Delaware Research Foundation and the John Templeton Foundation Award No. 63595. M.P. is supported by the Department of Energy under Grant No.~DE-SC0011842 at the University of Minnesota. The authors would like to thank C. Hall, D. Mckinsey, A. Hook, B. Safdi and R. Sundrum for useful discussions.

\bibliographystyle{apsrev4-2}
\bibliography{reference.bib}

\end{document}